\documentclass[12pt]{iopart}
\usepackage{graphicx}
\usepackage{bm}
\usepackage{epsfig}
\usepackage{psfrag}

\def\dsize{\displaystyle}
\def\E{\mathop{\rm E}\nolimits}

\def\Re{\mathop{\rm Re}\nolimits}
\def\Ra{\mathop{\rm Ra}\nolimits}

\def\Pr{\mathop{\rm Pr}\nolimits}

\begin{document}

\title[Kinetic energy cascades]{Kinetic energy cascades in quasi-geostrophic  convection in a spherical shell}

\author{Maxim  Reshetnyak$^1$, Pavel Hejda$^2$}
\address{$^1$ 
Institute of the Physics of the Earth, Russian Acad.~Sci, 123995 Moscow, Russia}
\ead{m.reshetnyak@gmail.com}
\address{$^2$ Institute of Geophysics, Academy of Sciences, 141 31 Prague, Czech Republic}
\ead{ph@ig.cas.cz}

\begin{abstract}
We consider triadic nonlinear interaction in the Navier-Stokes equation for quasi-geostrophic convection in a spherical shell. This approach helps understanding the origin of kinetic energy transport in the system and the particular scheme of mode interaction, as well as the locality of the energy transfer. The peculiarity of convection in the sphere, concerned with excitation of Rossby waves,  is considered. The obtained results are compared with our previous study in Cartesian geometry.
\end{abstract}

\submitto{\PS}
\maketitle

\section{Introduction}
\label{s1}
The redistribution of the physical fields is a quite widespread phenomenon in Nature. 
 If we consider convective systems with dissipation, there are then two major effects which work in the opposite directions: diffusion (may be turbulent), which leads
 to the homogenization of the fields, and nonlinear interactions, 
 which can produce large gradients  and, in fact,  cause the self-organization of the system.
The most impressive examples of such behavior  is the appearance of  coherent structures in  convective systems \cite{Tabel},  topological pumping \cite{Drobyshevski} and turbulent diamagnetism in the dynamo \cite{Zeldovich}.

 The considered effects demonstrate pumping of the fields in physical space to some domain of a scale smaller  
 than the main scale of the system. If the subdomain is quite small, then  it is clear that pumping should also be accompanied by the redistribution of the 
 fields in wave space, e.g., the energy fluxes between the different scales relevant to the  nonlinear
 process, $\alpha$-effect  and 
 separation in the scales of magnetic helicity \cite{BrSubr2005, HR10} in the mean-field dynamo theory \cite{Kr}, already mentioned coherent structures in convection.
 It is worth noting that fluxes exist in the wave space even in quasistationary states provided that
 the scales of   the energy injection into the system and  the energy dissipation are different, see the introduction to  the problem in
 \cite{Frisch} and the review of some recent results in \cite{Verma}.

In general, such fluxes can be defined  for  various physical quantities, and the direction of their propagation is prescribed  by  many factors:  by the nature of the considered quantity, the number of dimensions in the system and even the geometry of the system (as we shall see latter). Restricting our further research to  convection, we recall that for the 3D homogeneous isotropic  turbulence (in the absence of rotation) there is a direct energy cascade of the kinetic energy $E_K$ from the large scales  to the small scales, where dissipation takes place. The second invariant of the Navier-Stokes equation in 3D in the inviscid limit   is kinetic helicity $\chi$. However, for isotropic turbulence its mean value is zero and its mean flux is also zero.

The situation changes in 2D, where the cascade of kinetic energy is inverse: from the small to the large scales. 
Two-dimensional idealization was a useful approach for describing geophysical turbulence and was able to capture many
  important features of the flow. With regard to the more realistic models, one should consider quasi-geostrophic flows, which in view of their 
 properties are somewhere  inbetween 3D and 2D. In such a flow, one still has 3 dimensions, but the dependence  of  angular rotation  ($z$ coordinate) on 
 direction is degenerated, see \cite{Pedlosky}. This flow is known by its  structures elongated
  along $z$. The perpendicular scale of these structures is very small and defined by the value of the Ekman number $\E$. At the critical Rayleigh number, 
  the first columnar mode has horizontal and vertical scales  $ {\cal O}(E^{1/3}),\, {\cal O}(1)$, respectively \cite{Busse70}. Note that $\E\sim 10^{-15}$ in the Earth's liquid core, and the horizontal scale is extremely small. 
As follows from our previous study for the rotating  rectangular box heated from below with periodic boundary conditions in the horizontal plane  \cite{HR08}, these structures can supply kinetic energy in both directions: to the larger as well as to the smaller scales, which can be important for understanding the energy budget in the planetary cores and for constructing semi-empirical models of turbulence.  However, the solution of the same equations in plane geometry differs from that in the sphere. 

 In Cartesian  geometry, the increase of the  heat sources  
characterized by the Rayleigh number   leads to the gradual increase of the number of growing wave modes
with smaller and larger scales than the scale of the first mode.  However, in spherical geometry the increase of $\Ra$ excites Rossby waves. As a result the columns start to oscillate in the direction of the axis of rotation of the sphere, which leads to the rotation of columns around it. The direction of this rotation is defined by the slope of the outer boundary. For a  spherical shell it is prograde and for a concave surface it has the opposite direction \cite{Busse02}. The appearance 
of this rotation (even differential) corresponds  to 
the axi-symmetrical mode in spectral space. Thus the scenarios of energy transfer in the  plane and in spherical geometries can be different. This is the motivation of our present study. Here, following \cite{HR08}, we discuss the fluxes of kinetic energy  in the standard Bousinesq equations used in the geodynamo for the different magnitudes of $\Ra$ and compare the results with the Cartesian geometry survey.

\section{Equations}

  The thermal convection  process
driven by the flows of
 incompressible fluid ($\nabla\cdot{\bf V}=0$) 
 in the Boussinesq approximation in a spherical  shell $(r_{\rm ICB}\le
 r\le
r_{\rm CMB})$
  rotating with the angular velocity $\Omega$ in the $\it  z$-direction is described  by
the  Navier-Stokes equation \begin{equation}\dsize \Pr^{
-1}\E\left( {\partial{\bf V}\over\partial t}+\left({\bf
V}\cdot\nabla\right){\bf V}\right) = -\nabla P+ {\bf F}+
\E\nabla^2{\bf V} \label{Nav}
\end{equation} and 
the heat flux equation for temperature fluctuations  $T$
\begin{equation}\dsize
{\partial T\over\partial t}+\left({\bf V}\cdot\nabla\right) \left(T +T_\circ\right)=
\nabla^2T, \label{therm}
\end{equation}
where  $\dsize T_\circ={{{r_{\rm ICB}/ r} -r_{\rm CMB}}\over r_{\rm CMB}-r_{\rm ICB}}$ is the solution of the
 heat flux equation  with fixed 
 temperatures (1,\, 0) at the boundaries ($r_{\rm ICB},\,r_{\rm CMB} $ ) in the absence of convection. 
Hereinafter ICB denotes the inner core boundary, CMB the core mantle boundary
and $\left(r,\, \theta,\, \varphi \right)$ is the spherical system of coordinates. 

The equations are scaled with the outer radius of the shell $\rm L$,
which makes the dimensionless radius $r_{\rm CMB} = 1$; the inner core
radius $r_{\rm ICB}$ is equal to 0.35, which is the value valid
 for the Earth.
 Velocity $\bf V$,  pressure $P$ and the typical
diffusion time $t$ are measured in units of
$\rm \kappa/L$,  $\rm \rho\kappa^2/L^2$ and $\rm L^2/\kappa$, respectively, where $\kappa$ is the 
thermal  diffusivity, $\rho$ is density,  $\dsize \Pr={\nu\over
\kappa}$ is the Prandtl number,
  $\rm \dsize \E = {\nu\over 2\Omega L^2}$ is the
Ekman number and $\nu$ is the kinematic viscosity.

  Force $\bf F$
includes the Coriolis and Archimedean  effects:
\begin{equation}\dsize
{\bf F}= -{\bf 1}_z\times{\bf V}+ \Ra  Tr{\bf 1}_r,\label{force}
\end{equation}
where 
 $\bf 1_z$ is the unit vector along the axis of rotation,
$\dsize \Ra
={\alpha g_o\delta T {\rm L}\over  2\Omega\kappa}$ is the modified
Rayleigh number, $\alpha$ is the coefficient of volume expansion,
$\delta T$ is the unit of temperature,
 and $g_o$ is the gravitational acceleration at $r=r_{\rm CMB}$.

The inner core,  $r\le r_{\rm ICB}$, with surface $S_{\rm ICB}$,
can rotate around axis $\it z$ due to the viscous 
torque, caused by the  no-slip boundary conditions, used at both the boundaries $r_{\rm ICB},\,r_{\rm CMB} $.  The momentum equation for the angular velocity 
$\omega$ of the inner core has the form:
\begin{equation} \dsize \Pr^{-1} \,  I \, {\partial\omega\over\partial t} =
 \, r_{\rm ICB} \oint\limits_{S_{\rm ICB}}
{\partial
V_\varphi\over\partial r} \, \sin\theta \, \mathrm{d}S,\label{core}
\end{equation} where $I$ is 
the inner-core moment of inertia along the
${\it z}$-axis. 

Equations (\ref{Nav}--\ref{core}) were solved using the standard spherical functions decomposition  accompanied by a poloidal-toroidal decomposition of vector field $\bf V$. The fast Chebyshev transform was used in the $r$-direction. 
The mesh grid in physical space was $128^3$. The Fortran code was parallelized in the $r$-direction using MPI. The details of the spherical function and Chebyshev polynomial decomposition  can be  found in \cite{GR84, Tg99, Simit}.

\section{Quasi-geostrophic convection}
\begin{figure}[t]
\vskip -10.0cm
\begin{minipage}[t]{.35\linewidth}
\hskip 4cm
\includegraphics[width=15cm]{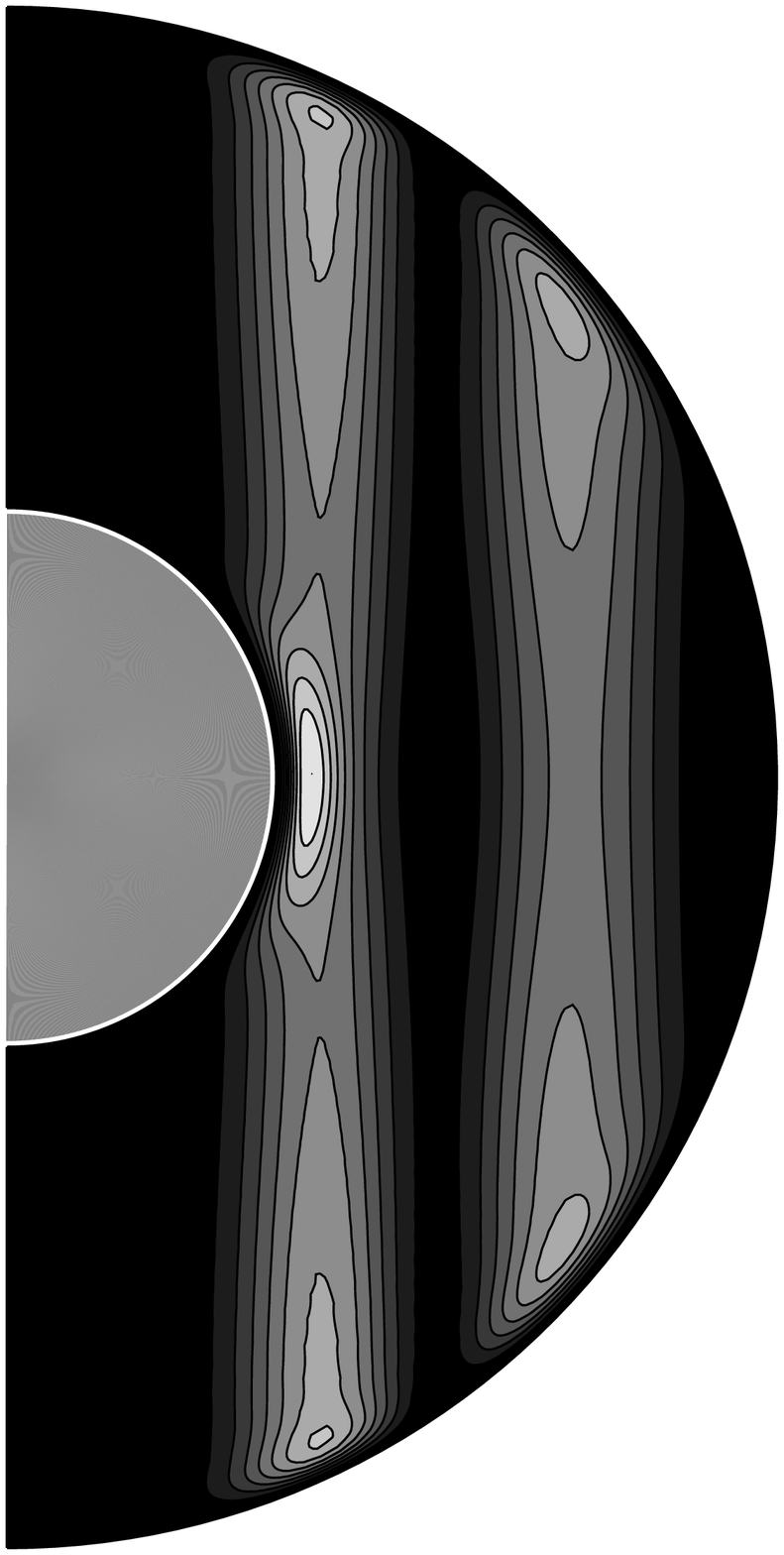}
\end{minipage}\hskip 4cm
\begin{minipage}[t]{.35\linewidth}
\includegraphics[width=15cm]{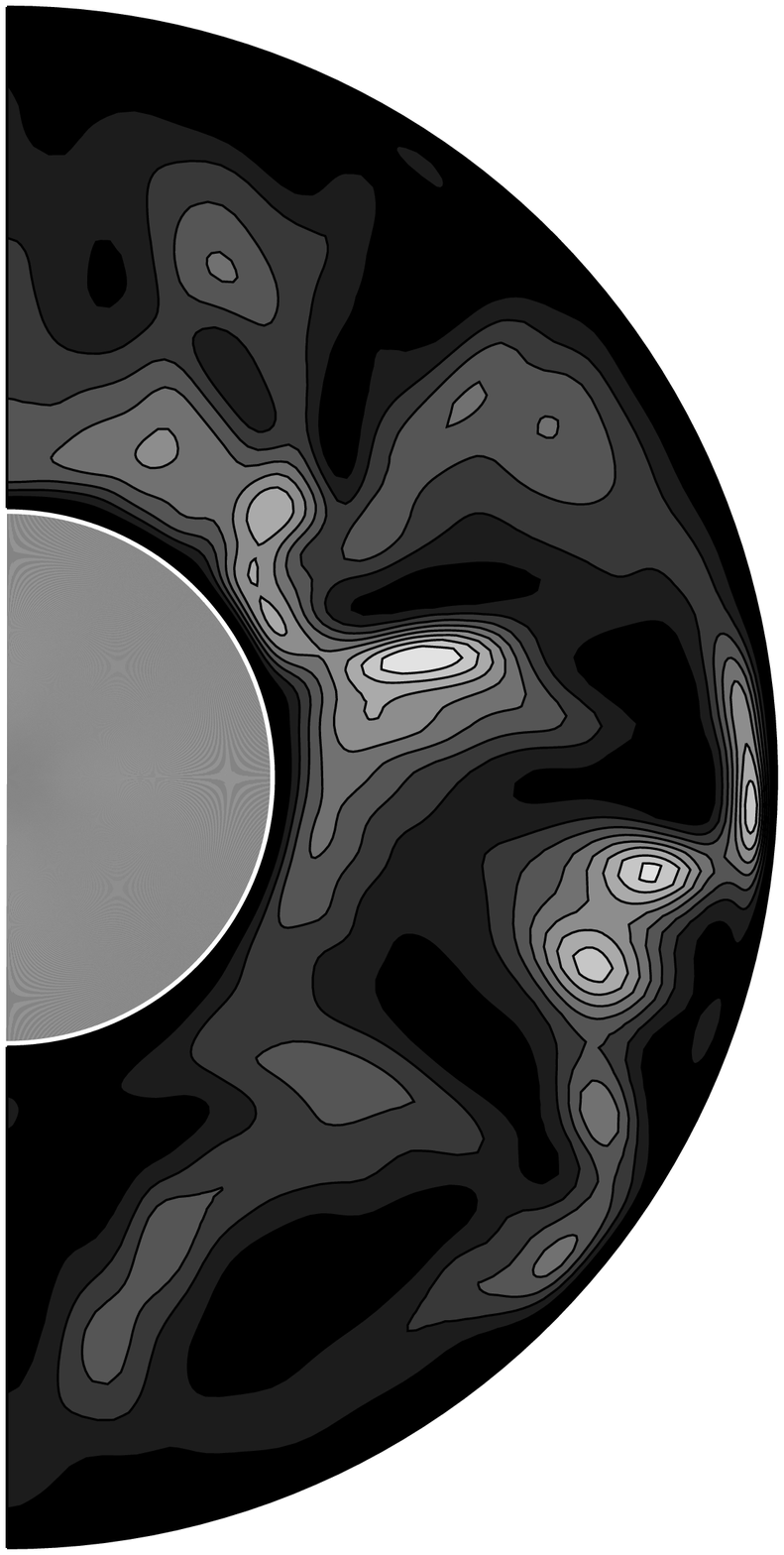}
\end{minipage}
\vskip -5cm \hskip 6cm\Large a \hskip 5cm b
\vskip 5.0cm
 \caption{
Meridional section of kinetic energy $E_K$ for   a) $\E=2\, 10^{-4}$, $\Pr=1$, $\Ra=40$,   $\Re\sim 17$, $(0,\, 60)$, with rotation, and 
b) $\E=1$, $\Pr=1$, $\Ra=2.5\, 10^6$,  
 $\Re\sim 3.4\, 10^2$, $(0,\, 5\, 10^4)$ without rotation (Coriolis force is switched off).
 Numbers in  parentheses correspond to the range of the field. } 
\label{fig1}
\end{figure}

\begin{figure}[t]
\vskip -10.0cm
\begin{minipage}[t]{.35\linewidth}
\hskip 4cm
\includegraphics[width=15cm]{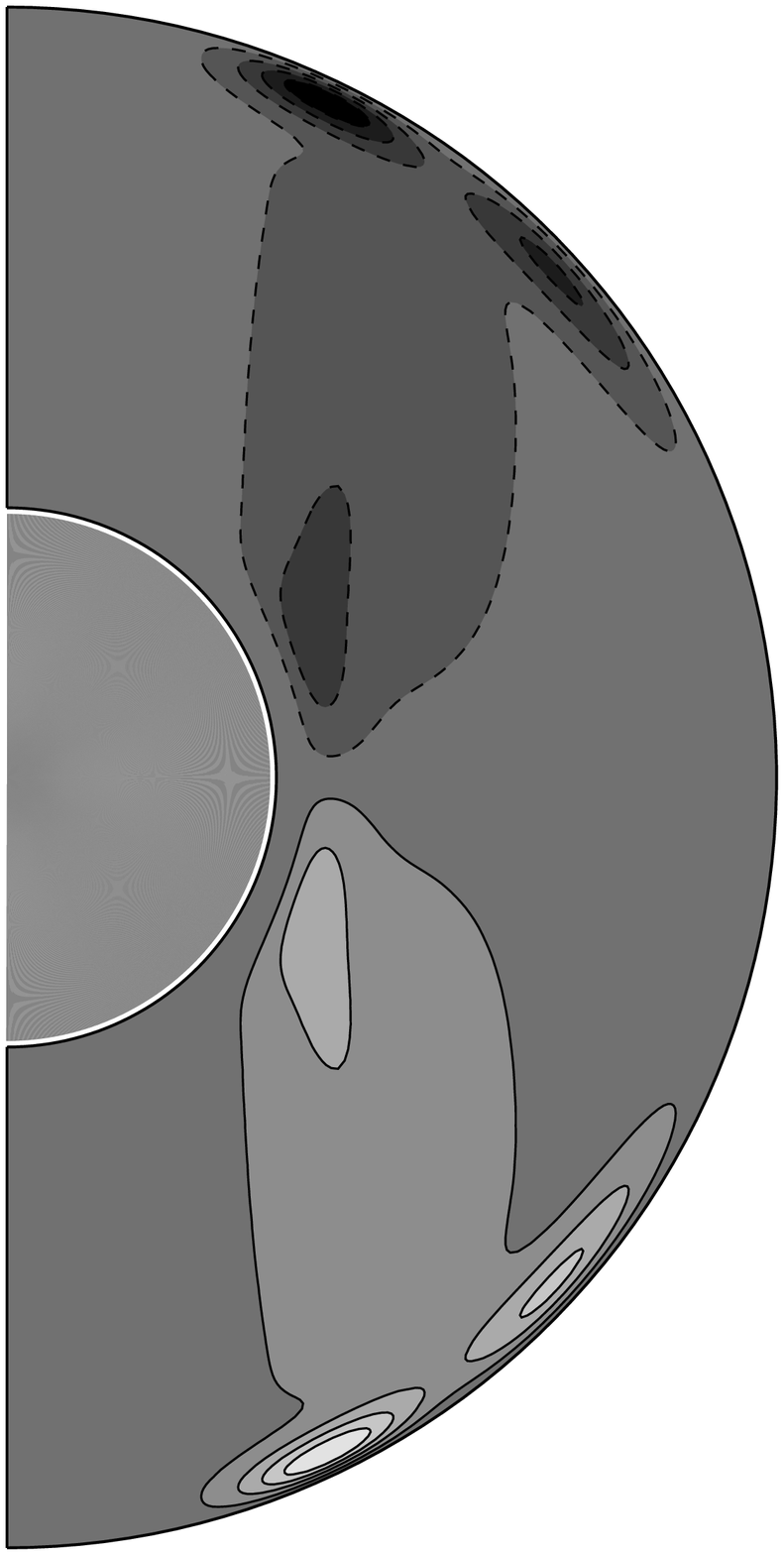}
\end{minipage}\hskip 4cm
\begin{minipage}[t]{.35\linewidth}
\includegraphics[width=15cm]{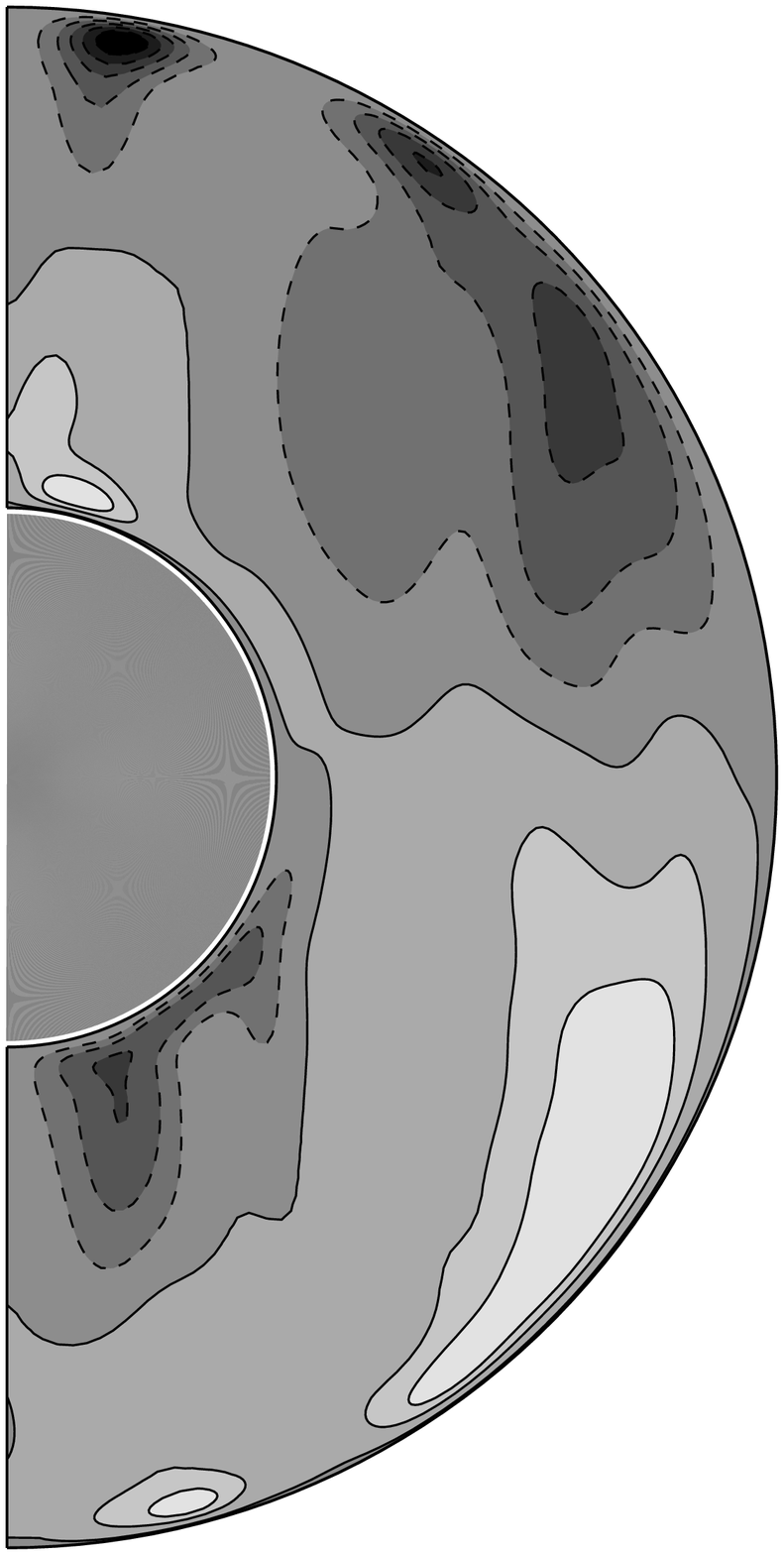}
\end{minipage}
\vskip -5cm \hskip 6cm\Large a \hskip 5cm b
\vskip 5.0cm
 \caption{
Meridional section of the mean kinetic helicity $\chi$ for  $\E=2\, 10^{-4}$, $\Pr=1$ a) $\Ra=40$,  $\Re\sim 17$, 
$(-1.2\, 10^3,\, 1.2\, 10^3)$ and b) $\Ra=4\, 10^2$, $\Re\sim 2\, 10^2$, 
 $(-3\, 10^5,\, 3\, 10^5)$.
The dotted isolines correspond to negative values. } 
\label{fig2}
\end{figure}

Rapid rotation is a quite familiar phenomenon in geophysics. A strong Coriolis force leads to the appearance  of 
elongated structures (columns) along the axis of rotation, in contrast to the cellular patterns in  the non-rotating regime, where  the spherically symmetrical (when the fields are averaged in time) buoyancy forces dominate, see Fig.~\ref{fig1}.  
 These columns with a horizontal scale $l_d\sim \E^{1/3}{\rm L}\ll 1$ rotate around their axes, so that  the net helicity 
 in the northern hemisphere for  small Rayleigh numbers is negative and in the southern hemisphere positive, see  Fig.~\ref{fig2}(a). 
Usually, the combination of the mean kinetic helicity $\chi=<{\bf V}\cdot {\rm rot}\, {\bf V}>$ (closely connected with the  $\alpha$-effect) with differential rotation is considered as an explanation of the existence of the large-scale planetary  magnetic fields. The helicity  depends on the regularity of the flow and ratio of the Coriolis and Archemedean forces: the increase of the heat sources 
 leads to the spread of convection to the Taylor cylinder (TC) accompanied by strong differential rotation. As a result, $\chi$ changes sign in the middle of the spherical shell in TC, being positive at the inner-core boundary (ICB) and negative at the core-mantle boundary (CMB) in the northern hemisphere (the helicity in TC still retains its dipole structure with respect to the equatorial plane). Before we refer the reader to  the more detailed study of  kinetic helicity, including the influence of the magnetic field on $\chi$  \cite{SD08}, let us stress that, from the point of view of the  mean-field theory \cite{Kr}, $\chi$ in TC and outside it are not the same: in TC $\chi$ is produced by the large-scale motions near the boundaries, and in the outer part it has a cyclonic nature with $l_d\ll 1$, so that the separation of the fields  into the large   and small  scales, adopted in the theory, is valid.

\section{Spectral properties}

\begin{figure}[t]
\vskip 0.0cm
\psfrag{m}{\Large $m+1$}
\begin{minipage}[t]{.45\linewidth}
\psfrag{D1}{\Large $E_K$}
\hskip 2cm 
\includegraphics[width=10cm]{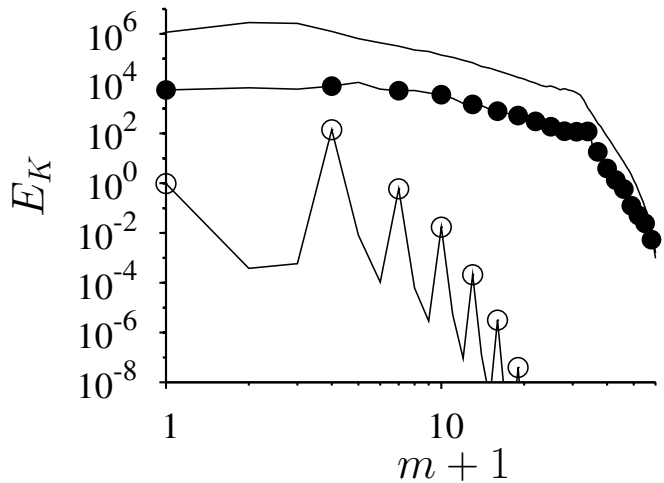}
\end{minipage}\hfill
 \caption{Spectra of kinetic energy  on the azimuthal wave number $m$ with rotation for $\Ra=40$ (open circles), $\Ra=800$ (solid circles)   and $\Ra=2.5\, 10^6$ (solid line) multiplied by $10^2$ without rotation. 
 To use the logarithmic scale in the $m$-direction,  the abscissa coordinate is shifted by 1. This means that the axi-symmetrical mode corresponds to value 1.}
\label{fig3}
\end{figure}

  As follows from the behavior of convection in physical space, rotation leads to the appearance of 
 the small scales in the horizontal plane. Hereinafter we consider the spectral 
properties in the azimuthal direction, bearing in mind that the  one-dimensional  spectrum 
$S(m)$ of field $F(r,\, \theta,\, \varphi)$ means:
$S(m)=\int\int f(m)\overline{f(m)}\, r^2  \sin\theta dr\, d\theta$, where $f$ is the Fourier transform of $F$ and 
 $\overline{f}$ is the complex conjugate of $f$. The spectra of  three  regimes are presented 
in Fig.~\ref{fig3}.
\newpage
 Near the onset of convection there is the well-pronounced peak in the spectra ($m=3$), corresponding to the horizontal scale of the column. Increase of $\Ra$ leads to the filling of the gaps in the spectra and transforming the saw-like structures to smooth curves. The slope of the spectrum for the developed rotating convection (solid circles) is less steep than for the non-rotating field because of the blocking of the nonlinear energy transfer over the spectrum by rotation.

To study this problem we recall the technique discussed in \cite{Frisch, Verma} and our recent papers 
\cite{HR08,  HR09}. The main idea is to allocate a sphere  of radius $M$ in Fourier  space and consider the evolution in time  of kinetic energy $E_K^<(M)$ inside the sphere taking into account the integral flux of the kinetic energy 
$\it \Pi(M)$ from  the  outer part of the sphere in wave space with $m>M$. The exact  form of the flux in physical space is 
$\it \Pi(M)=-\left[\left({\bf V}\cdot  \nabla\right)\, {\bf V}\right]\cdot {\bf V}^<$, where $\bf V^<$ denotes the filtered velocity field with all harmonics with $m>M$ equal to zero. 
 As follows from \cite{HR09}, in Cartesian geometry 
 the total flux $\Pi(k)$, where $k=|{\bf k_x} + {\bf  k_y} + {\bf  k_z}|$,  is mainly defined by the 
perpendicular contribution  $\Pi(k_\perp)$ with $k_\perp=|{\bf k_x} + {\bf  k_y}|$. This is the
  motivation to consider further only the $M$-dependency. 
In contrast to our previous study \cite{HR08, HR09} we do not proceed with local-flux functions 
$\dsize {\cal T}=-{\partial {\it \Pi} \over \partial M}$, because the spherical case exhibits more irregular behavior of the flux than the flat geometry. Note also that $\it \Pi$ includes  the local flux through a particular $M$, as well as the non-local contribution between the two subvolumes. 

\begin{figure}[th]
\vskip 0.0cm
\psfrag{m}{\large $M+1$}
\begin{minipage}[t]{.45\linewidth}
\psfrag{D1}{\large $\it \Pi_K$}
\hskip 2cm
\epsfig{figure=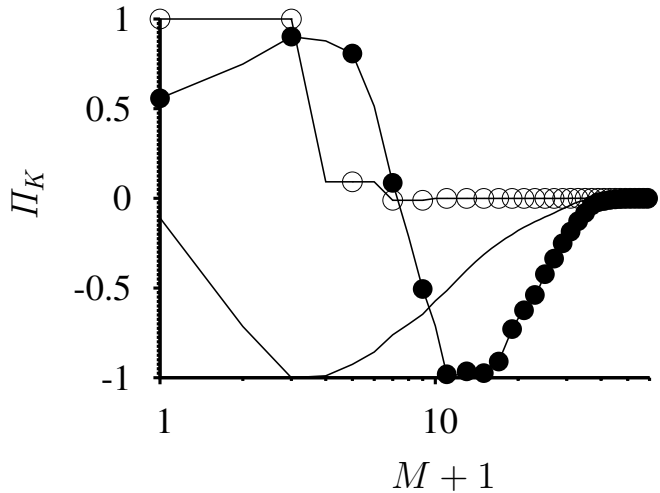,width=10cm}
\end{minipage}\hfill
\caption{Integral   flux $\it \Pi_K$ of kinetic energy as  a function of the azimuthal wavenumber for the three regimes with the same labels as in  Fig.\ref{fig3}.}
\label{fig4}
\end{figure}

 All the curves,
 see Fig.~\ref{fig4} for our regimes, 
 satisfy condition $\it \Pi=0$ for the large $M$ confirming the  conservation of kinetic energy in  the non-linear term of the Navier-Stokes equation. It is obvious  that rotation changes the energy transfer substantially. In the non-rotating regime, $\it \Pi$ is always negative, which means that the large scales feed the small scales (the direct cascade of energy). For the small $M$, the slope of increasing $ {\it \Pi}$ is constant, which means that the flux to the small scales   is non-local. The same situation holds for the larger $M$, where $\it \Pi$
  increases, but the slope  also remains constant. Due to this non-locality our solution differs from that for the Cartesian  geometry in \cite{HR08, HR09}.

For the geostrophic convection with small $\Ra$ we observe the inverse cascade of energy for $M=0-2$, $\it\Pi>0$. The break of the curve at $M=3$ means that, due to the sharp decrease of the spectrum, approximately the whole kinetic energy is in the sphere and $\it \Pi$ tends to zero. This is the reason why the direct cascade cannot be seen in  the plot for large $M$. 
 The increase of $\Ra$ decreases the relative flux to the axi-symmetrical mode twice. For larger $M$, $\it \Pi$ changes sign, which corresponds to the transition from the inverse cascade regime for $M<3$ to the direct cascade of the kinetic energy  for $M>3$.

\begin{figure}[t]
 \psfrag{D1}{\large $T_2$}
 \psfrag{m}{\large $K-Q$}
\vskip 0.0cm
\begin{minipage}[t]{.35\linewidth}
\hskip -0.0cm \epsfig{figure=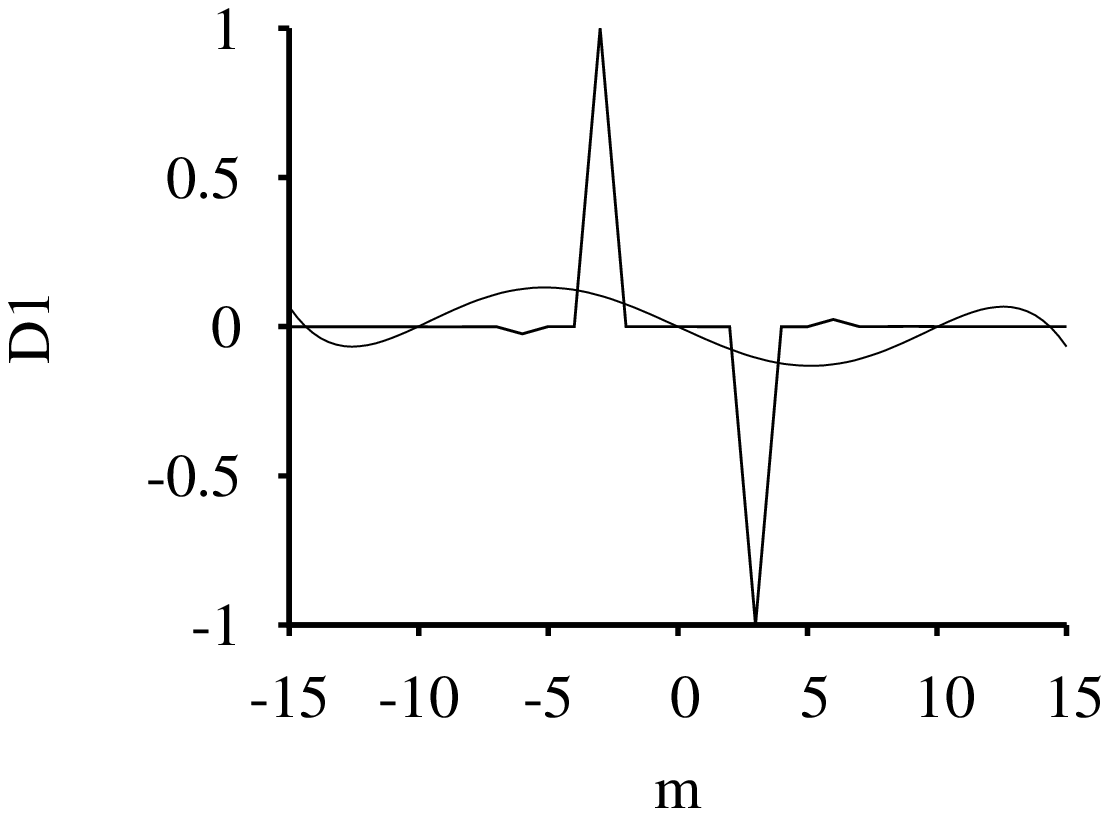,width=8cm}
\end{minipage}\hfill
\begin{minipage}[t]{.35\linewidth}
\hskip  -2.0cm \epsfig{figure=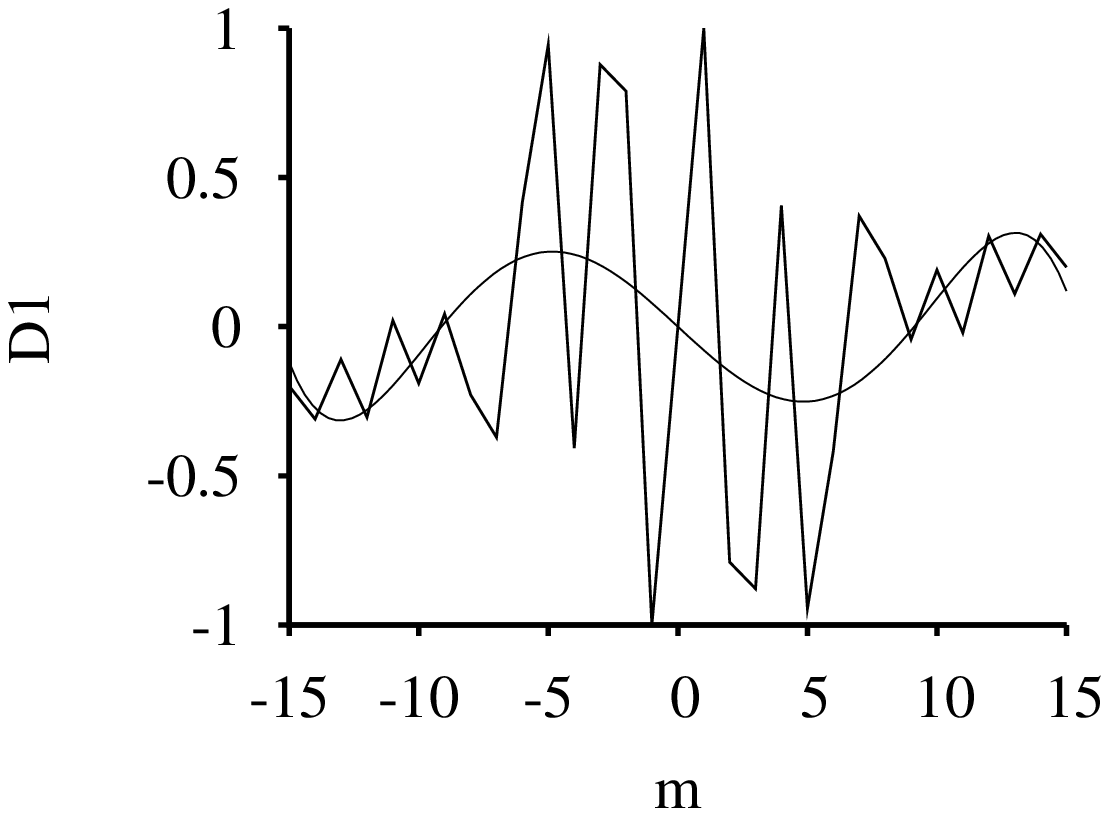,width=,width=8cm}
\end{minipage}\hfill
\vskip 0.0cm
\begin{minipage}[t]{.35\linewidth}
\hskip 4.0cm \epsfig{figure=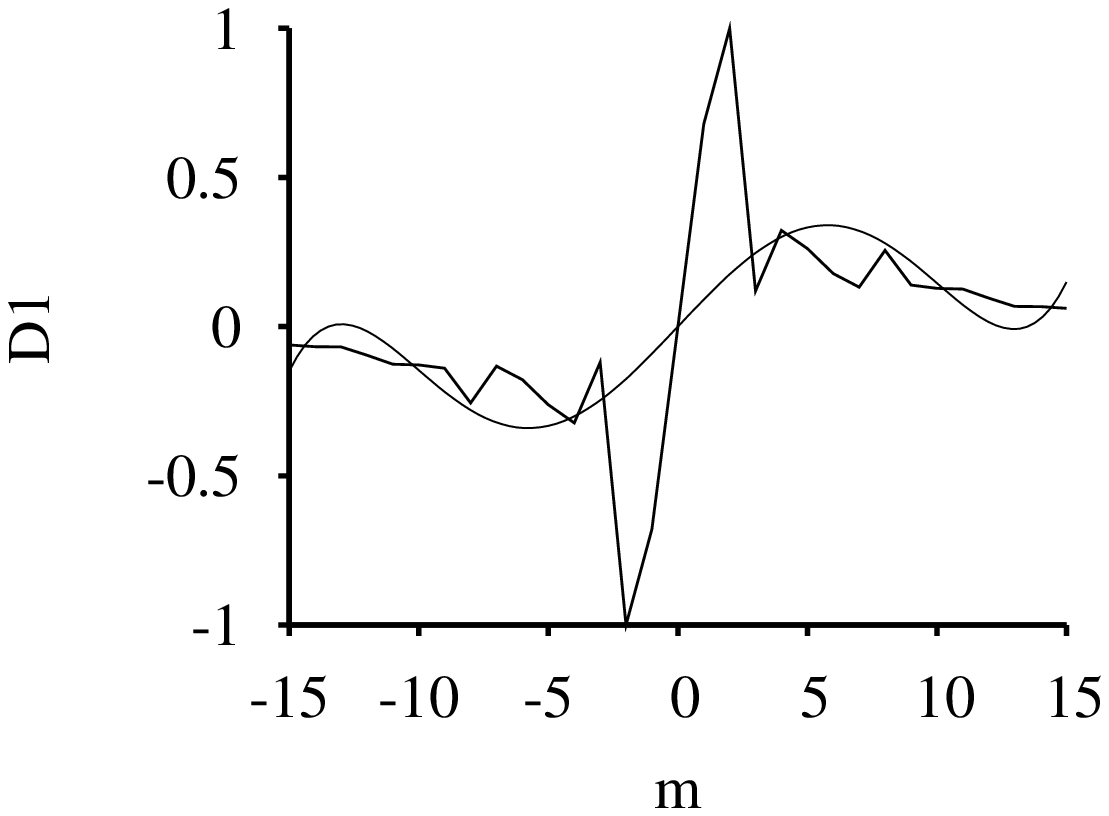,width=8cm}
\end{minipage}\hfill
\vskip -11.5cm \hskip 7.5cm\large a  \hskip 7.5cm  b \vskip 5cm \hskip 11cm c 
\vskip 5.5cm
 \caption{The fluxes of kinetic energy ${ T}_{2}(K-P)$  for the three regimes, 
see Fig.\ref{fig3},\ref{fig4}. Thin line corresponds to the  $5^{th}$ order polynomial approximation.}
\label{fig5}
\end{figure}

The rotation changes not only the direction of energy transfer over the spectrum, but also the structure of the triangle in wave space, when two modes with $m=P$ and $m=Q$ produce the third mode $m=K$.  To check this possibility, we have constructed the $T_2$ flux function which describes the input of energy of the $P$-mode to the $K$-mode: 
$\it T_2(P,K)=-\left[\left({\bf V}(P)\cdot  \nabla\right)\, {\bf V}\right]\cdot {\bf V}(K)$ \footnote{By definition $T_2$ is antisymmetric function.}.  It is convenient  to present $T_2$ as a function of $K-P$, see 
Fig.~\ref{fig5}.  There is clear evidence of the non-local transfer of energy between the modes for small $\Ra$ under rotating convection. The shift of the extrema from  the zero point describes the non-locality of interactions. This phenomenon can be explained as follows: the extrema correspond to the period ($m=3$) between the local maxima  in the spectra, see Fig.~\ref{fig3}. Negative $T_2$ for positive arguments means an inverse cascade. The increase of $\Ra$ leads to the complex interaction between the modes with both the direct and inverse cascades with a different level of non-locality. The approximation of the curve with polynomials demonstrates the evident inverse cascade and non-locality in interactions, see 
Fig.~\ref{fig5}. 
  The comparison of the non-local characteristics for the 
 spherical and  Cartesian geometries demonstrates the increase of the non-locality transfer in 
the former case.  There are many  reasons for this phenomenon. Due to the spherical boundary, the spectra on $m$ accumulate various scales in the $z$-directions.  
 The pure statistics in the spherical problem may be more important:  if  the number of 
 the columns is $\approx k_\perp^2$ in the flat geometry,  their number is much smaller in the spherical geometry, because they are mainly distributed near  TC. The other reason  is that considered in \cite{HR09}:   $k_\perp=8$ for the columns  is larger than that in the spherical geometry ($M=3$) and we stand far away from the similarity region, even if we have similar  resolutions in the models. That is why Cartesian geometry is so often used in MHD simulations to get scalings.

\section{Discussion}
We have tried to present the regimes of geostrophic convection, well-known in the geodynamo community, by traditional tools of the turbulent community. Even if the solution is quasi-stationary in physical space, there are non-zero fluxes of energy in wave space, proving that convection is not localized in wave space. 
 Our results demonstrate that recent simulations for the usually used grids do not exhibit local transfer of energy, even without rotation. Rotation brings inverse flux into the first harmonics. In this way we can consider the excitation of the axi-symmetrical rotation of the columns around geographical axes to be the result of the inverse cascade. This phenomenon has no analog in the case of  convection in the box, where rotation destroys rotating rolls with $m=0$, for details of the K\"uppers-Lortz instability refer to \cite{KL69, JR}. In some sense,  the periodical boundary condition used in the Cartesian geometry corresponds to the case with the zero curvature of the boundary considered in  \cite{Busse02}. 
 
As can be seen from the presented simulations, the relative  location of energy transfer  will increase with 
 increasing resolution of the model, but we believe that the mechanism of  excitation of the axi-symmetrical flow due to the non-local inverse energy transfer   will be the same. We  hope that these results will also be interesting for  solar modeling, where this technique could be used to check the energy exchange between the different radial layers and could thus help in constructing more realistic turbulent models.

\end{document}